\documentclass{JHEP3}

\usepackage{amsmath,amssymb,graphics}

\usepackage{epsfig,multicol}

\newcommand{\ba}{\begin{equation} \aligned}
\newcommand{\ea}{\endaligned \end{equation}}

\title{Spinor Field Realizations of the half-integer $W_{2,s}$ Strings}
\author{Shao-Wen Wei$^1$ ,
        Yu-Xiao Liu$^1$\footnote{Corresponding author.} ,
        Li-Jie Zhang$^2$ and Ji-Rong Ren$^1$ \\
    $^1$Institute of Theoretical Physics, Lanzhou University,
           Lanzhou 730000, P. R. China\\
    $^2$Department of Physics, Shanghai University,
           Shanghai 200444, P. R. China \\
    E-mail: \email{weishaow06@lzu.cn},
            \email{liuyx@lzu.edu.cn},
            \email{lijzhang@shu.edu.cn}}

\abstract{The grading Becchi-Rouet-Stora-Tyutin (BRST) method gives
a way to construct the integer $W_{2,s}$ strings, where the BRST
charge is written as $Q_B=Q_0+Q_1$. Using this method, we
reconstruct the nilpotent BRST charges $Q_{0}$ for the integer
$W_{2,s}$ strings and the half-integer $W_{2,s}$ strings. Then we
construct the exact grading BRST charge with spinor fields and give
the new realizations of the half-integer $W_{2,s}$ strings for the
cases of $s=3/2$, $5/2$, and $7/2$. }


\begin{document}

\section{Introduction}
\label{secIntroduction}

$W$ algebra \cite{Zamolodchikov1985,FateevNPB1987} was discovered
in two-dimensional field theories in the middle of the 1980's,
much work has been focused on the classification of it and on the
study of $W$ gravity and $W$ string theories. Furthermore, $W$
algebra plays a central role in many areas of two-dimensional
physics. It appears in the quantum Hall effect
\cite{KarabaliNPB1994} and black holes
\cite{IsoPRD2007,BonoraJHEP2008}, in lattice models of statistical
mechanics at criticality, and in other physical models
\cite{Leznov1989,FeherPR1992} and so on.

In all applications of $W$ algebra, the investigation of the $W$
strings is more interesting and important. The idea of building
$W$ string theories was first developed in Ref.
\cite{GervaisNPB1987}. Since $W_{3}$ is the simplest $W$ algebra,
most of the efforts in constructing $W$ string theories have been
concentrated on it
\cite{Thierry-MiegPLB1987,Bais1992,LuNPB1993,HornfeckPLB1993,OhtaPLB1995,Isaev2008,ZhuPZB1993,BergshoeffNPB1994}.
Using the method of Lagrangian Realization and Hamiltonian
Reduction, other strings and superstrings also had been studied in
\cite{BoerCMP1994,DeckmynPRD1995,BoreschNPB1995,MadsenCMP1997}.
Later Pope $et$ $al$. discovered that the
Becchi-Rouet-Stora-Tyutin (BRST \cite{BRST}) method provides the
only viable approach to the quantization of $W$ string theories.
Using a kind of canonical transformation \cite{LuNPB1993}, the
BRST operator could be written in the form of $Q_{B}=Q_{0}+Q_{1}$,
which is graded. Using this method, the scalar field realizations
of $W_{2,s}$ strings have been obtained for $s$=3,4,5,6,7
\cite{LuIJMPA1994,PopePLB1994}. Subsequently, the methods to
construct the spinor field realization of critical $W_{2,s}$
strings and $W_{N}$ strings were also found in Refs.
\cite{ZhaoPLB2000,WeiPLB2001}. Assuming the BRST charges of the
$W_{2,s}$ strings and $W_{N}$ strings are graded, the exact spinor
field realizations of $W_{2,s}(s=3,4,5,6)$ strings and
$W_{N}(N=4,5,6)$ strings were obtained
\cite{ZhaoPLB2000,WeiPLB2001,ZhangPRD2001,LiuCTP2004}. Recently,
we constructed the nilpotent BRST charges of the spinor
non-critical $W_{2,s}$ ($s$=3,4) strings by taking account of the
property of spinor field \cite{DuanNPB2004}. It was shown that
certain integer $W_{2,s}$ ($s$=3,4) algebras can be linearized as
the $W_{1,2,s}$ algebras by the inclusion of a spin-1 current.
This provides a way of obtaining new realizations of the integer
$W_{2,s}$ strings. Based on the linear $W_{1,2,s}$ algebras, the
ghost field realizations and the spinor realizations of the spinor
non-critical $W_{2,s}$ strings were given in Refs.
\cite{LiuJHEP2005,ZhangCTP2006}.

The half-integer $W_{2,s}$ algebra with $s=3/2$ is known as the
$N=1$ superconformal algebra or $SW(3/2,2)$ superconformal
algebra. Recently, several constructions of the algebra were
obtained in Refs. \cite{NoyvertJHEP2002,MichihiroJHEP2002} . Much
research had been done also on $N=2$ superconformal algebra, in
which another $s=3/2$ current is added
\cite{NobuyoshiPLB1994,PabloPLB1994,BellucciNPB1998,NathanPLB1994}.
Correspondingly, the $N=1$ and $N=2$ superstrings had been
investigated in \cite{WangNPB1992,KhviengiaNPB1995}. In addition,
the case of $s=5/2$ was investigated in Ref. \cite{XuiIJMP1995}.
It was found that this algebra can not be closed with spin-2
current $T$ and spin-5/2 current $G$ only, and so there is no
realization for this algebra. But by introducing other spin
currents, this algebra may be linearized \cite{BellucciPLB1995},
which implies that the realizations of the algebra could be found.
The $W_{2,s}~(s=7/2,9/2,11/2,13/2,15/2)$ algebras were also
studied in Ref. \cite{BlumenhagenNPB1991}, in which the values of
central charges for these algebras were given.

Since up to now there is no work focused on the research of spinor
field realizations of the half-integer $W_{2,s}$ strings, we will
construct the nilpotent BRST charges of spinor half-integer
$W_{2,s}$ strings for the first time by using the grading BRST
method. This method provides a more easy way to construct
$W_{2,s}$ algebras and strings. And the physical states can be
obtained by the grading form of BRST operators for these strings.
Firstly, we reconstruct the BRST charges $Q_{0}$ for integer
$W_{2,s}$ strings and the half-integer $W_{2,s}$ strings. For each
case we obtain four solutions. All these solutions indicate that
we can construct the BRST charges $Q_B$ for both the integer
$W_{2,s}$ strings and the half-integer $W_{2,s}$ strings by using
scalar field $\varphi$ together with spinor field $\psi$. But for
simplicity, we only consider the realizations with spinor field
$\psi$. With these solutions, we discuss the exact BRST charges
for $s={3}/{2},{5}/{2},{7}/{2}$ in detail and we then find that
many new valuable solutions will be obtained with increasing of
spin $s$. All these results will be of importance for embedding
the Virasoro string into the $W_{2,s}$ strings and the
half-integer $W_{2,s}$ strings, and they may provide the essential
ingredients to help us better understanding the fundamental
properties of the half-integer $W_{2,s}$ strings. Furthermore,
after giving the explicit realizations and BRST charges, the
physical states of these $W$ strings can be investigated.

This paper is organized as follows: in Section \ref{secReview} we
give a brief review of the grading BRST method for scalar field
and spinor field realizations of the integer $W_{2,s}$ strings. In
Section \ref{secQ0} we reconstruct the general BRST charges $Q_0$
for the integer and the half-integer $W_{2,s}$ strings. Using
these results, we construct the BRST charges $Q_B$ for
$W_{2,s}\;(s={3}/{2},{5}/{2},{7}/{2})$ in Section \ref{secW2s}.
Finally the conclusions are drawn in the last section.

\section{Review of the grading BRST method}
\label{secReview}

\indent In 1993, Pope {\em et. al.} \cite{LuNPB1993} discovered a
kind of canonical transformation that leads a considerable
simplification of the BRST operator and the physical states of
$W_{3}$ string. This is a nonlinear realization involving a scalar
field $\varphi$ whilst the spin-2 and spin-3 ghost fields are
introduced. Then the BRST operator could be written in the form of
$Q_{B}=Q_{0}+Q_{1}$, where $Q_{0}$ has grade $(1, 0)$ and $Q_{1}$
has grade $(0, 1)$, with $(p, q)$ denoting the grading of an
operator with ghost number $p$ for the redefined spin-2 $(b, c)$
ghost system and ghost number $q$ for the redefined spin-$s$
($\beta$, $\gamma$) ghost system. In particular $Q_{1}$ only
involves $\varphi$, $\beta$, $\gamma$. This leads to an immediate
generalization, the $W_{2,s}$ strings
\cite{PopePLB1994,PopeCQG1994}, whose BRST operator has the
similar form except that the ($\beta$,$\gamma$) system is a
spin-$s$ current rather than a spin-3 current.

Following Refs. \cite{PopePLB1994,PopeCQG1994}, the BRST operator
for the scalar field realizations of $W_{2,s}$ strings can be
written as
\begin{eqnarray}
    Q_{B}&=&Q_{0}+Q_{1}, \label{QB}     \\
    Q_{0}&=&\oint dz\;J_{0}=\oint dz\; cT, \label{Q0}   \\
    Q_{1}&=&\oint dz\;J_{1}=\oint dz\; \gamma F(\varphi,\beta,\gamma), \label{Q1}
\end{eqnarray}
where $J_{0}$, $J_{1}$ have spin 1 and $T$ are constructed as
\begin{equation}
    T=T^{eff}+T_{\varphi}+\frac{1}{2}T_{bc}+T_{\beta\gamma},
    \label{Tpope}
\end{equation}
here the energy-momentum tensors are given by
\begin{eqnarray}
 T_{\varphi} &=&-\frac{1}{2}(\partial\varphi)^{2}
               -\alpha\partial^{2}\varphi, \\
 T_{bc}&=&-2b\partial c-\partial b c, \\
 T_{\beta \gamma}&=&-s \beta \partial \gamma
               -(s-1)\partial \beta\gamma, \\
 T^{eff}&=&-\frac{1}{2} \eta_{\mu \nu}{\partial}X^{\mu}
               {\partial}X^{\nu}
               -i a_{\mu} \partial^{2} X^{\mu}.
\end{eqnarray}
The operator $F(\varphi,\beta,\gamma)$ has spin $s$ and ghost number
zero. Because of the grading character of $Q_{B}$, there exist the
nilpotency conditions \begin{equation}
Q_{0}^{2}=Q_{1}^{2}=\{Q_{0},Q_{1}\}=0,
\end{equation}
where the first nilpotency condition provides that the total central
charge vanishes, i.e.,
\begin{equation}
-26-2(6s^{2}-6s+1)+1+12\alpha^{2}+C^{eff}=0, \label{Ceff}
\end{equation}
the remaining two conditions determine the precise form of the
operator $F(\varphi,\beta,\gamma)$ appearing in Eq. (\ref{Q1}).

Recently, in Ref. \cite{ZhaoPLB2000}, the authors generalized the
scalar field grading BRST method above developed by Pope {\em et
al} to the spinor field realization of $W_{2,s}$ strings. The BRST
charge for the spinor field realization of $W_{2,s}$ strings takes
the similar form (\ref{QB})-(\ref{Q1}) of the scalar case but with
$F=F(\psi,\beta,\gamma)$, and the energy-momentum tensor $T$ was
constructed as
\begin{equation}
T= T^{eff}+T_{\psi}+KT_{bc}+yT_{\beta\gamma}, \label{Tweihao}
\end{equation}
in which $K$,$y$ are pending constants and the spinor field $\psi$
has spin 1/2 and satisfies the OPE
\begin{equation}
\psi (z) \psi (\omega)\sim -\frac{1}{z-\omega}.
\end{equation}

The study can be carried out by introducing the $(b, c)$ ghost
system for the spin-2 current, and the $(\beta, \gamma)$ ghost
system for the spin-$s$ current, where $b$ has spin 2 and $c$ has
spin $-1$ whilst $\beta$ has spin $s$ and $\gamma$ has spin
$(1-s)$. The ghost fields $b, c, \beta, \gamma$ are all bosonic
and commuting. The energy-momentum tensors in Eq. (\ref{Tweihao})
take the following forms
\begin{eqnarray}
T_{\psi}&=&-\frac{1}{2}\partial\psi\psi, \\
T_{bc}&=&2b\partial c+\partial bc, \\
T_{\beta\gamma}&=&s\beta\partial\gamma+(s-1)\partial\beta\gamma,\\
T^{eff}&=&-\frac{1}{2}\eta_{\mu\nu}\partial Y^{\mu} Y^{\nu}.
\end{eqnarray}
Using the properties of spinor field and noticing the multi-spinor
$Y^{\mu}$ is a multi-spinor, the BRST charge here is also graded
with $Q_{0}^{2}=Q_{1}^{2}=\{Q_{0},Q_{1}\}=0$. Different from the
scalar realizations, the first nilpotency condition $Q_{0}^{2}=0$
is satisfied for an arbitrary $s$, and there has no constraint on
the total central charge. The remaining two conditions determine
the precise form of the operator $F(\psi,\beta,\gamma)$ and the
exact $y$. The particular method used to construct
$F(\psi,\beta,\gamma)$ can be found in
\cite{ZhaoPLB2000,WeiPLB2001,ZhangPRD2001,LiuCTP2004}, in which
the solutions for $s=3,4,5,6$ have been obtained and the
discussions of any $s$ were carried out.

\section{Construct BRST charges $Q_{0}$ for the integer and the half-integer $W_{2,s}$ strings}
\label{secQ0}

After a brief review of the grading BRST method, we would like to
construct more general BRST charges $Q_{0}$ for the integer
$W_{2,s}$ strings and the half-integer $W_{2,s}$ strings.

Focusing on the form of $Q_0$ in Eq. (\ref{Q0}), here we write the
general form of the energy-momentum tensor $T$ as
\begin{equation}
 T=m_1 T^{eff}+m_2 T_{\psi}
 + m_3 T_{\varphi}+m_4 T_{bc}
 +m_5 T_{\beta\gamma},
 \label{Tgeneral}
\end{equation}
where $m_1-m_5$ are constants which should be determined by the
nilpotency of $Q_{0}$. The energy-momentum tensors on the right
hand side of (\ref{Tgeneral}) are
\begin{eqnarray}
T_{\psi}&=&-\frac{1}{2}\partial\psi\psi, \\
T_{\varphi}&=&-\frac{1}{2}(\partial \varphi)^{2}-\alpha
\partial^{2}\varphi, \\
T_{bc}&=&-2b\partial c-\partial bc, \\
T_{\beta\gamma}&=&-s\beta\partial\gamma-(s-1)\partial\beta\gamma,\\
T^{eff}&=&-\frac{1}{2} \eta_{\mu \nu}\partial X^{\mu}\partial
    X^{\nu}-i a_{\mu} \partial^{2} X^{\mu}-\frac{1}{2}\eta_{\mu\nu}\partial Y^{\mu}
    Y^{\nu},
\end{eqnarray}
here the $(b,c)$ ghost system corresponds to the spin-2 current,
and the $(\beta,\gamma)$ corresponds to the spin-$s$ one. $b$ has
spin 2 and $c$ has spin $-1$ whilst $\beta$ has spin $s$ and
$\gamma$ has spin $(1-s)$.

The spinor field $\psi$ has spin 1/2 and the scalar field $\varphi$
has spin 0, they satisfy the OPEs
 \begin{equation}
    \psi(z) \psi(\omega) \sim -\frac{1}{z-\omega}, \label{OPES}
\end{equation}
\begin{equation}
    \partial \varphi(z)
\partial\varphi(\omega) \sim -\frac{1}{(z-\omega)^{2}}.
\end{equation}
The OPE of $\psi$ with itself only has the first order pole and
$\partial\varphi$ only has the second order pole, they all just
contain one pole in their OPEs. The OPE of the effective
energy-momentum tensor $T^{eff}$ with itself is
 \begin{equation}
    T^{eff}(z) T^{eff}(\omega) \sim \frac{C^{eff}/2}{(z-\omega)^{4}}
+\frac{2T^{eff}}{(z-\omega)^{2}}+\frac{\partial T^{eff}}{z-\omega},
\end{equation}
where $C^{eff}$ is the central charge of $T^{eff}$. Different from
the cases of $\partial\varphi$ and $\psi$, one will find that this
OPE contains three poles, the first order pole, the second and the
forth ones, and the third order pole is vanished.

Next, we consider the nilpotency condition $Q_{0}^{2}=0$. Using
residue theorem, one may find this condition converts to require
the vanishing of the first order pole of $J_0(z) J_0(\omega)$.
Making use of the OPE relations, we could calculate the OPE of
$J_0(z) J_0(\omega)$, which contains three poles, but we are only
interesting in the first order pole, which has many terms. Then
let the coefficients of each terms to be zero will give constraint
equations to the parameters $m_1-m_5$. The number of these
equations may larger than that of the parameters, but these
equations are not linearly independent. After solving these
equations, we can determine the parameters $m_1-m_5$ and get the
solutions of energy-momentum tensor $T$. We find that the
nilpotency of $Q_0$ also requires the total central charge of $T$
to be zero for each solution.

\subsection{Solutions for the case $s =$ integer}

For the case $s=$ integer, the ghost fields $b,c,\beta,\gamma$ are
all fermionic and anticommuting. They satisfy the OPEs
\begin{equation}
 b(z) c(\omega) \sim \frac{1}{z-\omega}, \quad \beta(z) \gamma(\omega) \sim
    \frac{1}{z-\omega}. \label{OPEL}
\end{equation}
In other cases the OPEs vanish. We can see the OPEs of $b(z)
c(\omega)$ and $\beta(z) \gamma(\omega)$ only have first order
pole.

Using the OPEs (\ref{OPES})-(\ref{OPEL}) and considering the
condition $Q_{0}^{2}=0$, we obtain four solutions of $T$, and the
constraint on the total central charge for each solution:

\begin{itemize}

{\item Solution 1}

\begin{equation}
\begin{array}{c}
 T=T^{eff}+T_{\psi}+T_{\varphi}
   +\frac{1}{2}T_{bc}+T_{\beta\gamma}, \\
 -53+2C^{eff}+24\alpha^{2}+24s-24s^{2}=0. \label{t1}
\end{array}
\end{equation}

{\item Solution 2}

\begin{equation}
\begin{array}{c}
T=T^{eff}+T_{\varphi}+\frac{1}{2}T_{bc}+T_{\beta\gamma}, \\
-26-2(6s^{2}-6s+1)+1+12\alpha^{2}+C^{eff}=0.
\end{array}
\end{equation}

{\item Solution 3}

\begin{equation}
\begin{array}{c}
 T=T^{eff}+T_{\psi}+T_{\varphi}+\frac{1}{2}T_{bc}, \\
-49+2C^{eff}+24\alpha^{2}=0.
\end{array}
\end{equation}

{\item Solution 4}

\begin{equation}
\begin{array}{c}
T=T^{eff}+T_{\varphi}+\frac{1}{2}T_{bc}, \\
-25+C^{eff}+12\alpha^{2}=0.
\end{array}
\end{equation}

\end{itemize}

From above four solutions, we can see each solution contains the
energy-momentum tensors $T^{eff}$, $T_{bc}$ and $T_{\varphi}$. The
energy-momentum tensor $T_{\psi}$ is appeared only in solutions 1
and 3, and energy-momentum tensor $T_{\beta\gamma}$ only in
solutions 1 and 2. It is worth to note that solution 2 is in
accord with (\ref{Tpope}) and (\ref{Ceff}), which are the exact
expression of energy-momentum tensor and the constraint condition
on the central charge in \cite{LuIJMPA1994,PopePLB1994}, where
$W_{2,s}$ (s=3,4,5,6) strings were constructed with scalar field.
One may also find that solution 1 is the only one expression of
the energy-momentum tensor $T$ which contains five energy-momentum
tensors, and this may imply that we can give new realizations of
$W_{2,s}$ strings using scalar field $\varphi$ and spinor field
$\psi$. Choosing the expression (\ref{t1}) of energy-momentum
tensor $T$, we have constructed the $W_{2,3}$ string and found
that the result of $Q_{1}=\oint dz\; \gamma
F(\varphi,\psi,\beta,\gamma)$ does not contain the spinor field
$\psi$, and the result is exact the result in
\cite{LuIJMPA1994,PopePLB1994}. We will discuss the more general
case of $Q_{1}=\oint dz\; \gamma F(\varphi,\psi,b,c,\beta,\gamma)$
in our later work.

\subsection{Solutions for the case $s =$ half-integer}

Next, we will consider the half-integer $W_{2,s}$ strings. It is
important to note that for the half integer case, the ghost fields
$\beta$, $\gamma$ corresponding to spin-$s$ current are no more
fermionic and anticommuting but bosonic and commuting, they
satisfy the following OPE relation
\begin{equation}
\beta(z) \gamma(\omega) \sim -\frac{1}{z-\omega}.
\end{equation}

Following the procedure above, one can get four non-trivial
solutions also for the half-integer $W_{2,s}$ strings.

\begin{itemize}

{\item Solution 1}

\begin{equation}
\begin{array}{c}
T=T^{eff}+T_{\psi}+T_{\varphi}+\frac{1}{2}T_{bc}+T_{\beta\gamma}, \\
-51+2C^{eff}+4(1-6s+6s^{2})+2+24\alpha^{2}=0.
\end{array}
\end{equation}

{\item Solution 2}

\begin{equation}
\begin{array}{c}
T=T^{eff}+T_{\psi}+T_{\varphi}+\frac{1}{2}T_{bc}, \\
-51+2C^{eff}+2+24\alpha^{2}=0.
\end{array}
\end{equation}

{\item Solution 3}

\begin{equation}
\begin{array}{c}
  T=T^{eff}+T_{\psi}+\frac{1}{2}T_{bc}, \\
  -51+2C^{eff}=0. \\
\end{array}\label{T3}
\end{equation}

{\item Solution 4}
\begin{equation}
\begin{array}{c}
  T=T^{eff}+T_{\psi}+\frac{1}{2}T_{bc}+T_{\beta\gamma}, \\
  -51+2C^{eff}+4(1-6s+6s^{2})=0. \\
\end{array}\label{T4}
\end{equation}

\end{itemize}
One can see the energy-momentum tensors $T^{eff}$, $T_{\psi}$ and
$T_{bc}$ appear in each of these solutions. Energy-momentum tensor
$T_{\varphi}$ is contained in solutions 1 and 2 only, while
$T_{\beta \gamma}$ is contained in solutions 1 and 4 only. Here,
we need to point out that any realization of
$F(\psi,b,c,\beta,\gamma)$ for the half-integer $W_{2,s}$ strings
with $T$ given by solution 3 (or 4) is also a realization for the
case $T$ is given by solution 2 (or 1).

The total central charges of the matter sector for integer and the
half-integer $W_{2,s}$ strings are given in Table \ref{Tab:C1} and
Table \ref{Tab:C2} for each solution of $T$, respectively. It's
clear that, for solutions 1 and 2 of the integer $W_{2,s}$
strings, the total central charges of the matter sector for the
integer $W_{2,s}$ strings increase with spin $s$. But for
solutions 3 and 4, they are independent of $s$ and equal to 26,
which is resulted by the contribution of $T_{bc}$. Similarly, for
the half-integer $W_{2,s}$ strings, the central charges of the
matter sector in solutions 2 and 3 are independent of spin $s$.
But the central charge in solutions 1 and 4 are dependent of spin
$s$. When $s \geq 5/2$, the central charge of the matter sector
will be negative. However, we can substitute solution 1 for
solution 4 and set $\alpha \rightarrow i \alpha$, which has no
effect on the realizations of the half-integer $W_{2,s}$ strings.
Then by choosing the proper value of $\alpha$, the central charge
of the matter sector will be positive.

\begin{table}[h]
\begin{center}
\renewcommand\arraystretch{1.4}
\begin{tabular}{|c|c|c|c|c|c|}
  \hline
  ~ & $W_{2,3}$ & $W_{2,4}$ & $W_{2,5}$ & $W_{2,6}$ & $W_{2,7}$ \\
  \hline
  Solutions 1,2 & 100 & 172 & 268 & 388 & 532 \\
  \hline
  Solutions 3,4 & 26 & 26 & 26 & 26 & 26 \\
  \hline
\end{tabular}
\caption{The total central charges of the matter sector for the
integer $W_{2,s}$ strings \label{Tab:C1}}
\end{center}
\end{table}

\begin{table}[h]
\begin{center}
\renewcommand\arraystretch{1.4}
\begin{tabular}{|c|c|c|c|c|c|}
  \hline
  ~ & $W_{2,3/2}$ & $W_{2,5/2}$ & $W_{2,7/2}$ & $W_{2,9/2}$ & $W_{2,11/2}$ \\
  \hline
  Solutions 1,4 & 15 & $-$21 & $-$81 & $-$165 & $-$273 \\
  \hline
  Solutions 2,3 & 26 & 26 & 26 & 26 & 26 \\
  \hline
\end{tabular}
\caption{The total central charge of the matter sector for the
half-integer $W_{2,s}$ strings \label{Tab:C2}}
\end{center}
\end{table}

From above discussion, it implies that we can construct the
half-integer $W_{2,s}$ strings using scalar field $\varphi$ and
spinor field $\psi$. But for simplicity, in the next section, with
those solutions (\ref{T3}) and (\ref{T4}) for the energy-momentum
tensor $T$, we will construct the BRST charge for the half-integer
$W_{2,s}$ strings using spinor field $\psi$ only.

\section{Spinor field realizations of the half-integer
$W_{2,s}$ strings} \label{secW2s}

In present section, we will construct the spinor field
realizations of the half-integer $W_{2,s}$ strings for the case of
$s={3}/{2}$, ${5}/{2}$ and ${7}/{2}$ by introducing the fermionic
$(b_{1}, c_{1})$ ghost system for the spin-2 current. The OPE of
$b_{1}(z) c_{1}(\omega)$ is the same as (\ref{OPEL}). The BRST
charge $Q_B$ is also graded $Q_B=Q_0+Q_1$ with $Q_0$ taking the
form of (\ref{Q0}), while $T$ is given in (\ref{T3}) or
(\ref{T4}), and $Q_1$ is given by
\begin{eqnarray}
    Q_{1}&=&\oint dz\; \gamma F(\psi,b_{1},c_{1},\beta,\gamma).
    \label{Q1_new}
\end{eqnarray}
Using the grading BRST method, we will construct
$F(\psi,b_{1},c_{1},\beta,\gamma)$ and discuss the spinor field
realizations of the half-integer $W_{2,s}$ strings for the case of
$s={3}/{2}$, ${5}/{2}$ and ${7}/{2}$.

The general procedure is described as follows. First, write down
all possible terms of $F$ with $\psi,b_{1},c_{1},\beta,\gamma$ by
considering the spin `$s$' of each term and ghost number zero.
Then leave out all the total differential terms in $\gamma F$
since their contribution to $Q_{1}$ is zero. Next, get the
coefficient equations by using the nilpotency conditions
$Q_{1}^{2}=\{Q_{0},Q_{1}\}=0$. Finally, determine the coefficients
in $F$ by solving these equations.


\subsection{Spinor field realizations of the $W_{2,{3}/{2}}$ string}

In this case, the most extensive combinations of
$F(\psi,b_{1},c_{1},\beta,\gamma)$ in (\ref{Q1_new}) with correct
spin and ghost number can be constructed as follows
\begin{equation} \aligned
 F(\psi,b_{1},c_{1},\beta,\gamma) =& f_{1} \partial\psi
            +f_{2} \psi b_{1} c_{1}
            +f_{3} \psi \beta \gamma
            +f_{4} \partial \beta c_{1} \\
            &+f_{5} \beta \partial c_{1}
            +f_{6} b_{1} \gamma
            + f_{7} c_{1} \beta^{2} \gamma. \label{W3/2}
\endaligned \end{equation}

Substituting (\ref{W3/2}) back into (\ref{Q1_new}) and imposing the
nilpotency conditions $Q_{1}^{2}=\{Q_{0},Q_{1}\}=0$, we can
determine $f_{i}$ (i=1,2,...,7).

\textbf{Case 1: $T=T^{eff}+T_{\psi}+\frac{1}{2}T_{bc}$}

For the case, the solution is
\begin{equation*}
  f_{1} =  f_{2} =  f_{3} = f_{4} = f_{7} = 0,
\end{equation*}
and the remaining $f_{5}$, $f_{6}$ are arbitrary constants but do
not vanish at the same time.

\textbf{Case 2: $T=T^{eff}+T_{\psi}
+\frac{1}{2}T_{bc}+T_{\beta\gamma}$}

For the case we can just get
\begin{equation*}
  f_{1} =  f_{2} =  f_{3} = f_{4} = f_{5} = f_{6} = f_{7} = 0,
\end{equation*}
this is a trivial result that all the coefficients of the terms in
$F(\psi,b_{1},c_{1},\beta,\gamma)$ are vanished.\\

\subsection{Spinor field realizations of the $W_{2,{5}/{2}}$ string}

Similarly, for the case s=${5}/{2}$,
$F(\psi,b_{1},c_{1},\beta,\gamma)$ is expressed in the following
form
\begin{equation} \aligned
F(\psi,b_{1},c_{1},\beta,\gamma) & = g_{1} \partial^{2}\psi +g_{2}
\partial \psi b_{1} c_{1}+g_{3} \psi \partial \beta \gamma
+g_{4} \psi \beta \partial \gamma+g_{5} \psi \partial b_{1} c_{1}  \\
                &  +g_{6} \psi b_{1} \partial c_{1}+ g_{7} \psi b_{1} c_{1} \beta \gamma
                +g_{8}\psi \partial b_{1} b_{1} \gamma^{2}
+g_{9} \partial b_{1} b_{1} c_{1} \gamma +g_{10} b_{1} \beta
\partial \gamma \gamma \\ &  +g_{11} b_{1} \beta^{2} \gamma^{3}+ g_{12}
\partial b_{1} \beta \gamma^{2}
                +g_{13} b_{1} \partial^{2} \gamma
+g_{14} \partial b_{1} \partial \gamma +g_{15} c_{1} \partial \beta  \\
                &  +g_{16} \partial c_{1} \beta
                + g_{17} c_{1} \beta^{2} \gamma.
\endaligned \end{equation}

It is clear that the number of these terms in
$F(\psi,b_{1},c_{1},\beta,\gamma)$ for $s=5/2$ is larger than
$s=3/2$, and this may give more solutions.

For the first case $T=T^{eff}+T_{\psi}+\frac{1}{2}T_{bc}$, there
are seven solutions:

\begin{itemize}

{\item Solution 1}
\begin{equation*}
 g_{i}  = 0 \quad (i=1-12,15-17),
\end{equation*}
and $g_{13}$ and $g_{14}$ are arbitrary constants but do not
vanish at the same time.

{\item Solution 2}
\begin{eqnarray*}
 g_{i} &=& 0 \quad (i=1-8,11,14-17),\\
 g_{9} &=& g_{10} =3M_{1},\;\;\;
 g_{12} = M_{1},\;\;\;
 g_{13} = -9 M_{1},
\end{eqnarray*}
where $M_{1}$ is a non-zero constant.

{\item Solution 3}
\begin{eqnarray*}
 g_{i} & =& 0 \quad (i=1-7,11,15-17),\\
 g_8 &=& M_2,\;\;\;
 g_{9} = -36 M_3,\;\;\;
 g_{10} =-36 M_3,\;\;\;\\
 g_{12} &=& 16 M_3,\;\;\;
 g_{13} = -270 M_3,\;\;\;
 g_{14} = -{135} M_3,
\end{eqnarray*}
where $M_2$, $M_3$ are arbitrary constants but do not vanish at the
same time.

{\item Solution 4}
\begin{eqnarray*}
 g_{i} & =& 0 \quad (i=1-7,15-17),\\
 g_8 &=& M_4,\;\;\;
 g_{9} = -24 M_5,\;\;\;
 g_{10} = -60 M_5,\;\;\; \\
 g_{11} &=& 4 M_5,\;\;\;
 g_{12} = -8 M_5,\;\;\;
 g_{13} = {33} M_5,\;\;\;
 g_{14} = -{39} M_5,
\end{eqnarray*}
where $M_4$, $M_5$ are arbitrary constants but do not vanish at the
same time.

{\item Solution 5}
\begin{eqnarray*}
 g_{i} & =& 0 \quad (i=1-7,9,15-17),\\
 g_{8} &=& M_6,\;\;\;
 g_{10} = -12 M_7,\;\;\;
 g_{11} = M_7,\;\;\;  \\
 g_{12} &=& -2 M_7,\;\;\;
 g_{13} = -6 M_7,\;\;\;
 g_{14} = -6 M_7,
\end{eqnarray*}
where $M_6$ and $M_7$ are arbitrary constants but do not vanish at
the same time.

{\item Solution 6}
\begin{eqnarray*}
 g_{i} & =& 0 \quad (i=1-9,11,13,15,17),\\
 g_{10} &=& 33 M_8,\;\;\;
 g_{12} =4 M_8, \;\;\;
 g_{14} = -18 M_8, \;\;\;
 g_{16} = M_9, \;\;\;
\end{eqnarray*}
where $M_8$ and $M_9$ are arbitrary constants but do not vanish at
the same time.

{\item Solution 7}
\begin{eqnarray*}
 g_{i} & =& 0 \quad (i=1-8,15-17), \\
 g_{9} &=& -24 M_{10},\;\;\;
 g_{10} = -60 M_{10},\;\;\;
 g_{11} =4 M_{10},\;\;\; \\
 g_{12} &=& -8 M_{10},\;\;\;
 g_{13} ={33} M_{10},\;\;\;
 g_{14} = -{39} M_{10},
\end{eqnarray*}
where $M_{10}$ is a non-zero constant.
\end{itemize}

It is worth pointing out that every term in each of these solutions
does not contain spinor field $\psi$, but we will show that, for
$s=7/2$, this situation will be changed.

For the second case, i.e.,
$T=T^{eff}+T_{\psi}+\frac{1}{2}T_{bc}+T_{\beta\gamma}$, we get
\begin{equation*}
  f_{i} = 0 \quad (i=1-17).
\end{equation*}
Like the case of $W_{2,{3}/{2}}$, this solution is also trivial.

\subsection{Spinor field realizations of the $W_{2,{7}/{2}}$ string}

For $s={7}/{2}$, $F(\psi,b_{1},c_{1},\beta,\gamma)$ can be
expressed in the following form:
\begin{equation} \aligned
 &F(\psi,b_{1},c_{1},\beta,\gamma) \\
 & = h_{1} \partial^{3}\psi
+h_{2}
\partial^{2} \psi b_{1} c_{1}+h_{3} \partial^{2} \psi \partial b_{1}
b_{1} \partial \gamma \gamma +h_{4} \partial \psi \beta^{2}
\gamma^{2}+h_{5} \partial \psi
\partial b_{1} c_{1}
+h_{6} \partial \psi b_{1} \partial c_{1}\\
                & + h_{7} \partial \psi \beta \partial \gamma  +h_{8} \partial \psi \partial^{2} b_{1} b_{1}
                \partial \gamma \gamma
+h_{9} \partial \psi \partial b_{1} b_{1} \partial^{2} \gamma
\gamma+h_{10} \partial \psi \partial b_{1} b_{1} \partial \gamma
\partial \gamma  \\ & +h_{11} \psi \beta^{3} \gamma^{3}+ h_{12}
\psi \partial^{2} b_{1} c_{1}
               +h_{13} \psi \partial b_{1} \partial c_{1} +h_{14}
\psi b_{1} \partial^{2} c_{1}+h_{15} \psi \partial^{2} \beta \gamma
+h_{16} \psi \beta
\partial^{2} \gamma  \\ & +h_{17} \psi \partial \beta \partial \gamma
+h_{18} \psi \partial^{4} b_{1} b_{1} \gamma^{2}+ h_{19} \psi
\partial^{3} b_{1} \partial b_{1} \gamma^{2}
 +h_{20}\psi \partial^{3} b_{1} b_{1} \partial \gamma \gamma +h_{21} \psi \partial^{2} b_{1} \partial b_{1} \partial
\gamma \gamma \\ &+h_{22} \psi \partial b_{1} b_{1} \partial^{3}
\gamma \gamma +h_{23}\psi \partial b_{1} b_{1} \partial^{2} \gamma
\partial \gamma+ h_{24} \psi \partial b_{1} c_{1} \beta \gamma
+h_{25} \psi b_{1} \partial c_{1} \beta \gamma +h_{26} \psi b_{1}
c_{1} \partial \beta \gamma \\  &+h_{27} \psi b_{1} c_{1} \beta
\partial \gamma +h_{28} \partial \psi \psi \partial^{2} b_{1} \gamma+h_{29} \partial \psi \psi b_{1} \partial^{2} \gamma
+h_{30} \partial \psi \psi \partial b_{1} \partial \gamma + h_{31}
c_{1} \beta^{2} \gamma \\ & +h_{32} c_{1}  \partial \beta +h_{33}
\partial c_{1} \beta +h_{34} \partial^{2} b_{1} b_{1} c_{1} \beta
\gamma^{2} +h_{35} \partial b_{1} b_{1} c_{1} \beta \partial \gamma
\gamma+ h_{36} \partial^{3} b_{1} \partial \gamma \\ & +h_{37}
\partial^{2} b_{2} \partial^{2} \gamma +h_{38}
\partial b_{1} \partial^{3} \gamma+h_{39} b_{1} \partial^{4} \gamma
+h_{40}\partial^{3} b_{1} \partial^{2} b_{1} b_{1} \gamma^{3}+h_{41}
\partial^{3} b_{1} \partial b_{1} b_{1} \partial \gamma \gamma^{2}
\\ & +h_{42} \partial^{2} b_{1} b_{1} c_{1} \partial \gamma+ h_{43}
\partial^{2} b_{1} \beta \partial\gamma \gamma +h_{44} \partial^{2}
b_{1} \partial b_{1} b_{1} \partial \gamma \partial \gamma \gamma
+h_{45}b_{1} \beta \partial^{3} \gamma \gamma \\ & +h_{46}
\partial^{2} b_{1} \partial b_{1} c_{1} \gamma+h_{47} \partial^{2} b_{1} \partial b_{1} b_{1} \partial^{2} \gamma \gamma^{2}
+ h_{48} \partial b_{1} b_{1} \partial c_{1} \partial \gamma+h_{49}
\partial b_{1} \partial \beta \partial \gamma \gamma \\ & +h_{50}
\partial b_{1} \beta \partial \gamma \partial \gamma+h_{51}
\partial b_{1} b_{1} c_{1} \partial ^{2} \gamma+h_{52} b_{1} \partial \beta \partial ^{2} \gamma \gamma
+h_{53} b_{1} \beta \partial ^{2} \gamma \partial \gamma+h_{54}
\partial b_{1} \partial^{2} \beta \gamma^{2} \\ & + h_{55} b_{1} \partial^{2} \beta \partial \gamma \gamma
+h_{56} b_{1} \partial \beta \partial \gamma \partial \gamma+h_{57}
\partial^{2} b_{1} b_{1} \partial c_{1} \gamma +h_{58} \partial b_{1} b_{1} \partial^{2} c_{1} \gamma
+h_{59} b_{1} \partial^{3} \beta \gamma^{2} \endaligned
\end{equation}

For the first case $T=T^{eff}+T_{\psi}+\frac{1}{2}T_{bc}$, one
will get three solutions

\begin{itemize}

{\item Solution 1}
\begin{eqnarray*}
 h_{i} &=& 0 \quad (i=1,2,4-7,11-17,19,21-59),\\
 h_{3} &=&  M_{11},\;\;\;
 h_{8} =  M_{11},\;\;\;
 h_{9} = M_{11},\;\;\; \\
 h_{10} &=& 2 M_{11},\;\;\;
 h_{18} = M_{12},\;\;\;
 h_{20} = 3 M_{12},
\end{eqnarray*}
where $M_{11}$ and $M_{12}$ are arbitrary constants but do not
vanish at the same time. This case is different from those above
since it contains the spinor field $\psi$ at each of the exist
terms.

{\item Solution 2}
\begin{eqnarray*}
 h_{i} &=& 0 \quad (i=1-35,40-48,50,51,53,57,58),\\
 h_{49} &=& M_{13}-3 M_{14},\;\;\;
 h_{52} = M_{13}-3 M_{14},\;\;\;
 h_{54} = M_{14},\;\;\; \\
 h_{55} &=& M_{13},\;\;\;
 h_{56} = 2 M_{13}-6 M_{14},\;\;\;
 h_{59} = M_{14},
\end{eqnarray*}
where $M_{13},M_{14},M_{15}$ and $h_{i}(i=36-39)$ are arbitrary
constants but do not vanish at the same time.

{\item Solution 3}
\begin{eqnarray*}
 h_{i} & =& 0 \quad (i=1-35,39-42,44-48,50,51,53,57,58),\\
 h_{38} & =& 2 M_{16},\;\;\;
 h_{54} =  2 M_{17},\;\;\;
 h_{55} =  M_{18},\;\;\;
 h_{59} =  2 M_{19},\;\;\;\\
 h_{36} &=& 2 M_{16}-9 M_{17}+9 M_{19},\;\;\;
 h_{37} = 3 M_{16},\;\;\;
 h_{43} = 6 (M_{19}-M_{17}),\;\;\;\\
 h_{49} &=& 6 M_{17}+M_{18}-12 M_{19},\;\;\;
 h_{52} = M_{18}-6 M_{19},\;\;\;
 h_{56} = 2 (M_{18}-6 M_{19}),
\end{eqnarray*}
where $M_{16},M_{17},M_{18},M_{19}$ are arbitrary constants but do
not vanish at the same time.

\end{itemize}

For the second case, i.e.,
$T=T^{eff}+T_{\psi}+\frac{1}{2}T_{bc}+T_{\beta\gamma}$, we get
\begin{eqnarray*}
 h_{i} & =& 0 \quad (i=1-2,4-7,11-17,19,21-59),\\
 h_{3} &=& M_{20},\;\;\;
 h_{8} = M_{20},\;\;\;
 h_{9} = M_{20},\;\;\; \\
 h_{10} &=& 2 M_{20},\;\;\;
 h_{18} = 2 M_{21},\;\;\;
 h_{20} = 2 M_{22},
\end{eqnarray*}
where $M_{20},M_{21},M_{22}$ are arbitrary constants but do not
vanish at the same time. This is a non-trivial case and it is
different from that of $W_{2,{3}/{2}}$ and $W_{2,{5}/{2}}$. One of
the main consequence of the solution may be that with increasing
of spin $s$, the number of those terms constructing
$F(\psi,b_{1},c_{1},\beta,\gamma)$ becomes large and may give more
new solutions.

Until now, we have constructed the explicit forms of
$F(\psi,b_{1},c_{1},\beta,\gamma)$ for $s=3/2$, $5/2$, $7/2$, and
find that with increasing of spin $s$, the results become more
interesting and complicated.

\section{Conclusion}
\label{secConclusion}

In this paper, we first give a brief review of the grading BRST
method to construct $W_{2,s}$ strings using scalar field $\varphi$
and spinor field $\psi$, respectively. Then we reconstruct the
BRST charge $Q_{0}$ more generally for both the integer $W_{2,s}$
strings and the half-integer $W_{2,s}$ strings. Each of them gives
four solutions, and each solution has the condition that the total
central charge must vanish. We find that for the half-integer
$W_{2,s}$ strings, when $s\geq 5/2$, the central charge of the
matter sector will be negative, but this can be solved by setting
$\alpha \rightarrow i\alpha$, which has no effect on the
realizations of the half-integer $W_{2,s}$ strings. We also find
that the energy-momentum tensor used in Refs.
\cite{LuIJMPA1994,PopePLB1994} is one of the special case of our
results. Based our results, we construct the $W_{2,3}$ strings
using scalar field $\varphi$ together with spinor field $\psi$,
but we find the results are the same as the ones obtained in Refs.
\cite{LuIJMPA1994,PopePLB1994}. But with these results, following
the procedure expressed detailedly in Section \ref{secW2s}, we
construct the nilpotent BRST charges $Q_B$ for the half-integer
$W_{2,s}$ strings for $s=3/2,5/2,7/2$ using spinor field $\psi$
only for the first time. When the spin $s$ takes $3/2$ and $5/2$,
each solution of $F(\psi,b_1,c_1,\beta,\gamma)$ for the case
$T=T^{eff}+T_{\psi}+\frac{1}{2}T_{bc}$ has not any term which
contains the spinor field $\psi$, and for the case
$T=T^{eff}+T_{\psi}+\frac{1}{2}T_{bc}+T_{\beta\gamma}$, there only
exists the trivial solution $F(\psi,b_{1},c_{1},\beta,\gamma)=0$.
But for $s=7/2$, some valuable  solutions are obtained. For the
case of $T=T^{eff}+T_{\psi}+\frac{1}{2}T_{bc}$, there exists one
solution in which each term of it contains the spinor field
$\psi$, and for the case of
$T=T^{eff}+T_{\psi}+\frac{1}{2}T_{bc}+T_{\beta\gamma}$, only one
non-trivial solution is found. These two solutions are different
from the cases of $s=3/2$ and $5/2$. This shows that with
increasing of spin $s$, many new valuable solutions should be
found. All these results obtained in this paper will be of
importance for embedding of the Virasoro string and superstrings
into the integer $W_{2,s}$ strings and the half-integer $W_{2,s}$
strings, and they may provide the essential ingredients to help us
better understanding the fundamental properties of the
half-integer $W_{2,s}$ strings. By giving the explicit
realizations and BRST charges, the study on the physical states of
these $W$ strings will become possible.

\section*{Acknowledgements}

It is a pleasure to thank Dr. Ji-Biao Wang for useful discussions.
This work was supported by the National Natural Science Foundation
of China (No. 10705013), the Doctor Education Fund of Educational
Department of China (No. 20070730055) and the Fundamental Research
Fund for Physics and Mathematics of Lanzhou University (No.
Lzu07002). L.J. Z acknowledges financial support from Shanghai
Education Commission.

\end{document}